\documentclass{PoS}

\title{Anomaly at finite density and chiral fermions on lattice
\vskip-3.4cm\hfill {\normalfont\large TIFR/TH/09-35}\vskip 3.1cm
}

\ShortTitle{Anomaly at finite density and chiral fermions on lattice}

\author{\speaker{Rajiv V. Gavai}  \\
        Department of Theoretical Physics, Tata Institute of Fundamental
        Research, \\ Homi Bhabha Road, Mumbai 400005, India.\\
        E-mail: \email{gavai@tifr.res.in}}

\author{Sayantan Sharma\thanks{Supported by the Shyamaprasad Mukherjee Fellowship
of CSIR, India.}\\
        Department of Theoretical Physics, Tata Institute of Fundamental
        Research,\\ Homi Bhabha Road, Mumbai 400005, India \\
        E-mail:\email{ssharma@theory.tifr.res.in}}

\abstract{Using both perturbation theory in the Euclidean formalism as well
as the non-perturbative Fujikawa's method, we verify that the chiral
anomaly equation remains unaffected in continuum QCD in the presence of
nonzero chemical potential, $\mu$.  We extend our considerations to
lattice fermions with exact chiral symmetry and discuss the
consequences for the recent Bloch-Wettig proposal for the Dirac operator at
finite chemical potential.  We propose a new simpler method of
incorporating $\mu$.  }

\FullConference{The XXVII International Symposium on Lattice Field Theory\\
		 July 26-31, 2009\\
		 Peking University, Beijing, China}

\begin{document}

\section{Introduction}

Since there are an equal number of {\em left}-handed and {\em right}-handed
particles in the lattice fermion propagator for each set of quantum
numbers, there is no chiral anomaly on the lattice for the na{\i}ve
fermions.  Canonical local formulations, such as the Wilson fermions, the
Kogut-Susskind fermions, the twisted mass fermions or the Creutz-Bori\c{c}i
fermions break the flavour singlet axial U(1) and some or all of the flavour
non-singlet axial symmetries. The overlap fermions, on the other hand,
preserve all chiral symmetries and have an index theorem as well.

Interestingly, chiral anomaly plays a crucial role in deciding the
structure of the QCD phase diagram for our world with two light quarks and one
moderately heavy quark.   A fundamental aspect of the QCD phase diagram
may be the existence of a critical point in the $T$-$\mu_B$ plane, where
$\mu_B$ denotes the baryonic chemical potential. The critical point is
expected on the basis of chiral symmetries and model considerations.  Since
we employ lattice techniques to investigate the phase diagram, the
presence on those chiral symmetries on the lattice is desired, if not
required, in the entire $T$-$\mu_B$ plane.  The popular staggered fermions
have a remnant chiral symmetry on the lattice but a not so well-defined
flavour number; two flavours are simulated using a square root of the
fermion determinant for them.   Clearly, use of overlap fermions with exact
chiral symmetry on lattice is desirable.

Introduction of chemical potential for the overlap fermions turns out to be
nontrivial.  Ideally, one should construct the conserved charge, $N$, as a
first step and then simply add $\mu N$  to the action.  For the local
fermions, this leads to $a^{-2}$ divergences in the continuum limit in
thermodynamic quantities such as the energy density or the number density.
One uses the freedom of adding irrelevant terms to cure them by multiplying
gauge links in positive/negative time direction by $\exp(a\mu)$ and $
\exp(- a\mu)$ respectively or more general functions which satisfy certain
conditions \cite{latmu}.  Note that this modification, indeed even the
simple addition of $\mu N$ does not change the chiral invariance of the
lattice action in anyway.  The non-local nature of the overlap fermions
makes this construction difficult, possibly even non-unique \cite{mand}.
It was proposed by Bloch and Wettig \cite{wettig} that the same
prescription of multiplying the gauge links, as above, may be used.  This
required them to extend the definition of the sign function employed in the
overlap operator since the argument of the sign function, namely the
Wilson-Dirac matrix, was no more $\gamma_5$-hermitian.  

Banerjee, Gavai and Sharma \cite{bgs} showed analytically that this
prescription does not have any $a^{-2}$ divergences in the continuum limit
but unfortunately the resultant overlap fermion action has no chiral
invariance on the lattice for nonzero $\mu$.  Of course, it is restored in
the continuum limit but so it is for the simpler local Wilson fermions: in
both cases the bare parameters of the action such as the quark mass or the
chemical potential will get renormalized and depend on the coupling $g$.
Alternatively, one can modify the chiral transformation, again by terms
which vanish in the continuum limit, to restore the chiral invariance.
This explains the results obtained by Bloch and Wettig that one even has an 
index theorem for the $\mu$-dependent overlap Dirac operator: the anomaly
equation itself is $\mu$-dependent \cite{wettig}.  Our work presented in this
talk was sparked by the curiosity about the fate of the anomaly relation in
the continuum on the introduction of nonzero chemical potential.

Before presenting our work let us remark, however, that the change in the
chiral transformation on the lattice implied by the Bloch-Wettig proposal
is undesirable since it makes the generators of the transformation to be
non-hermitian.  Worse still, the symmetry group itself changes for each and
every change of $\mu$, making this avenue not very useful for the original
problem of investigation of the QCD phase diagram in the $T$-$\mu_B$ plane.
In order to appreciate the loss, recall that the symmetry group remains the
{\em same} at each $T$ for $\mu=0$.  Therefore, a study of the temperature
dependence of the order parameter $\langle \bar \psi \psi \rangle$, and any
abrupt change in it at the transition temperature, can reveal the chiral
symmetry restoring transition, if any, and thereby the change in the nature
of the vacuum.  Were the symmetry group to change as a function of $T$, it
would be impossible to attribute a vanishing chiral condensate uniquely as
a change in the vacuum or a phase change, thereby eliminating the role of
the chiral condensate as the order parameter.  Thus one can indeed use the
freedom of adding terms irrelevant in the continuum limit to modify the
chiral transformation on the lattice but such a change should not rob one off 
the usual order parameter for studying the chiral transition.

\section{Anomaly at $T=0$ and $\mu\neq0$ in continuum}

Classically $\psi' = \exp {(i \alpha \gamma_5)}~ \psi$ and $\bar \psi' =
\bar \psi~ \exp {(i \alpha \gamma_5)} $ is a symmetry for the QCD action
for massless quarks , leading to the current conservation equation
$\partial_\mu J^\mu_5 = 0$.  At finite temperature and/or density, this
classical symmetry remains intact.  Quantum loop effects can, and do, cause
corrections already at $T=\mu=0$, leading to the anomaly equation. One
needs to compute $\langle \partial_\mu J^\mu_5 \rangle$ to check this.  The
famous calculation of the Adler-Bell-Jackiw(ABJ) triangle diagram for the
$U(1)$ case \cite{anom} demonstrated perturbatively that the axial
$U(1)$ is broken by quantum effects while Fujikawa \cite{fujikawa} provided
a new insight by showing that the anomaly arises due to the change of the
fermion measure under the corresponding transformation of the fermion
fields in the path integral method.  We have used both the methods to check
the fate of the anomaly equation at finite density, i.e, on introduction of
a nonzero chemical potential $\mu$ at zero temperature.

\subsection{Perturbative calculation}

The lowest order diagram is the canonincal ABJ axial
vector-vector-vector(AVV) triangle diagram. It is well-known that the
higher order diagrams do not contribute to the anomaly equation  at zero
density, neither do other diagrams like the square and pentagon diagrams.
We too therefore compute only AVV triangle diagram at finite density.  Our
notation for the QCD Lagrangian in the Euclidean space with the finite
number density term is the same as in \cite{kapusta}.  In order to maintain
consistency with the lattice literature, we have, however, chosen the Dirac
gamma matrices to be Hermitian, leading to the action, 

\begin{equation} \label{eqn:qcdl} 
\mathcal{L}=-\bar \psi({\not} D+m)\psi-\frac{1}{2}{\rm Tr ~}F_{\alpha\beta}
F_{\alpha\beta} +\mu\bar \psi\gamma_4 \psi~,  
\end{equation} 
where ${\not}D=\gamma_{\nu}(\partial_{\nu}-igA^{a}_{\nu}T_{a})$ with
$T_{a}$ being the generators of the SU(3) gauge group. The
$\gamma_5=\gamma_1 \gamma_2 \gamma_3 \gamma_4$ is also hermitian in our
case.  Using the canonical Euclidean space Feynman rules, the amplitude of
these AVV triangle diagrams can be computed. 
Denoting by $\Delta^{\lambda\rho\sigma}(k_1,k_2)$ the total amplitude and
contracting it with $q_{\lambda}$,  the $\langle \partial_\mu
j_\mu^5\rangle$ for $\mu \ne 0$ can be obtained from the triangle diagrams
for $\mu \ne 0$:
\begin{eqnarray}
\nonumber 
q_{\lambda}\Delta^{\lambda\rho\sigma}  &=&
  - i~g^2 {\rm tr}[T^a T^b]\int \frac{d^4 p}{(2\pi)^4}{\rm Tr ~}\left[\gamma^5
\frac{1}{{\not} p-{\not}q-i\mu\gamma^4}
\gamma^\sigma\frac{1}{{\not} p-{\not}k_1 -i\mu\gamma^4}\gamma^\rho
\right.\\ \nonumber
&-& \gamma^5\frac{1}{{\not} p-i\mu\gamma^4}
\gamma^\sigma\frac{1}{{\not} p-{\not}k_1-i\mu\gamma^4}\gamma^\rho
+ \gamma^5 \frac{1}{{\not} p-{\not}q-i\mu\gamma^4}
\gamma^\rho\frac{1}{{\not} p-{\not}k_2-i\mu\gamma^4}\gamma^\sigma \\
\label{eq:twoI}
&-& \left.  \gamma^5\frac{1}{{\not} p-i\mu\gamma^4}
\gamma^\rho\frac{1}{{\not} p-{\not}k_2-i\mu\gamma^4}\gamma^\sigma\right]~.
\end{eqnarray}

An inspection of the eq.(\ref{eq:twoI}) reveals quadratically divergent
integrals which differ from the usual case by merely having $\mu \ne 0$. The
computation therefore can be done the same way by  writing \cite{OurAno}
$q_{\lambda}\Delta^{\lambda\rho \sigma}=(-i)~{\rm ~tr~}[T^a T^b]g^2\int
\frac{d^4 p}{(2\pi)^4}\left[f(p-k_1,k_2) -f(p,k_2)+f(p-k_2,k_1)-f(p,k_1)
\right]$ for a suitably chosen function $f$ and introducing a
gauge-invariant cut-off.  The nonzero $\mu$ appears in $f$ in the
denominator as $[(p_4-i\mu)^2 +\vec p^2]$  with  linear in $\mu$ and a
$\mu$-independent term in the numerator; the $\mu^2$ term vanishes being
proportional to  Tr ~$[\gamma^5\gamma^4\gamma^\sigma\gamma^4 \gamma^\rho]
\sim \epsilon^{4\sigma 4 \rho}$. The final result  turns \cite{OurAno} out
to be the same anomaly relation as for $\mu$ = 0, since the $\mu$-dependent
terms appear with $\Lambda^{-1}$, and vanish as the cut-off $\Lambda \to
\infty$. The same computation is easily generalized \cite{OurAno} to nonzero 
temperatures.

\subsection{Non-perturbative calculation}

Fujikawa \cite{fujikawa} taught us how to compute the chiral anomaly
non-perturbatively using the path integral formalism.  Under the chiral
transformation of the fermion fields, the measure changes as 
\begin{equation}
 \mathcal{D}\bar \psi^{'}~\mathcal{D}\psi^{'}=\mathcal{D}\bar \psi~\mathcal{D}\psi
{\rm ~ Det}\vert\frac{\partial(\bar \psi^{'},\psi^{'})}{\partial(\bar \psi,\psi)}\vert
= \mathcal{D}\bar \psi~\mathcal{D}\psi~
 \exp(-2i\alpha \int d^4x~ {\rm Tr}\gamma_5 )~.
\label{eq:meas}
\end{equation}
The trace can be computed using the the eigenvectors of the operator
${\not}D$. It is an anti-Hermitian operator for $\mu =0$ with purely
imaginary eigenvalues and  the corresponding eigenvectors form  a complete
orthonormal basis. Using the fact $\{\gamma_5,{\not}D\}=0$, it is easy to
show\cite{OurAno} that Tr~$\gamma_5$ = 0 for the space of eigenvectors
with nonzero eigenvalues, leading to the usual Tr~$\gamma_5 = n_+ -n_-$
relation.

${\not}D(\mu)$ still anti-commutes with $\gamma_5$ but has  both an 
anti-Hermitian and a Hermitian term. Remarkably, it turns out \cite{OurAno} 
still to be diagonalizable with right and left eigenvectors which together
form a complete set using which essentially the same argument as above
leads to the same  Tr~$\gamma_5 = n_+ -n_-$ for $\mu \ne 0$ as well.  Note
that the zero modes are still defined with respect to ${\not}D(\mu = 0)$.
This should be contrasted with the index theorem for the overlap operator
proposed by Bloch-Wettig where the zero modes are  $\mu$-{\em dependent} 
\cite{wettig}.
This is perhaps natural since the corresponding chiral transformation is 
$\mu$-{\em dependent}, as discussed above.

Curiously, a ``gauge-like'' symmetry, defined by a {\em non-unitary}
transformation of the fermion fields, given by $\psi^{'}(\mathbf x,
\tau)=\exp{(\mu\tau)}~ \psi ( \mathbf x , \tau),~~ \bar \psi^{'}( \mathbf
x, \tau)= \bar \psi( \mathbf x, \tau)~\exp{(-\mu\tau)}~$, eliminates all
the $\mu$-dependent terms of the QCD action.  This transformation commutes
with the chiral transformations, explaining thus the preservation of the
same anomaly relation for $\mu \ne 0$.  If one were to insist on this
symmetry on the lattice, then addition of a simple $\mu N$ is easily seen
to be forbidden for any {\em local} fermion action.  Moreover, it can be
easily implemented for any such local action, and leads {\em naturally} to
all the known proposals \cite{latmu} of introduction of the chemical
potential on the lattice so far. 

While the symmetry clearly arises in the continuum due to the local nature
of the fermion action, one may choose to demand it for even non-local
lattice actions.  A natural proposal then for the overlap operator for nonzero
$\mu$ is $D_{ov}(\mu) = \exp{(-\mu\tau)}~D_{ov} (\mu =0)\exp{(\mu\tau)}$.
While this can be shown to i) satisfy the Ginsparg-Wilson relation
\cite{gw} and ii) lead divergence-free ideal gas results in the continuum
limit, it clearly does not commute with the nonlocal chiral transformations
unless they are made $\mu$-dependent too.  Its main advantage though could
still be that the usual sign function is used in this overlap operator.

\section{Two simple ideas for the lattice QCD at finite density}

\begin{figure}[htb]
\includegraphics[scale=0.6]{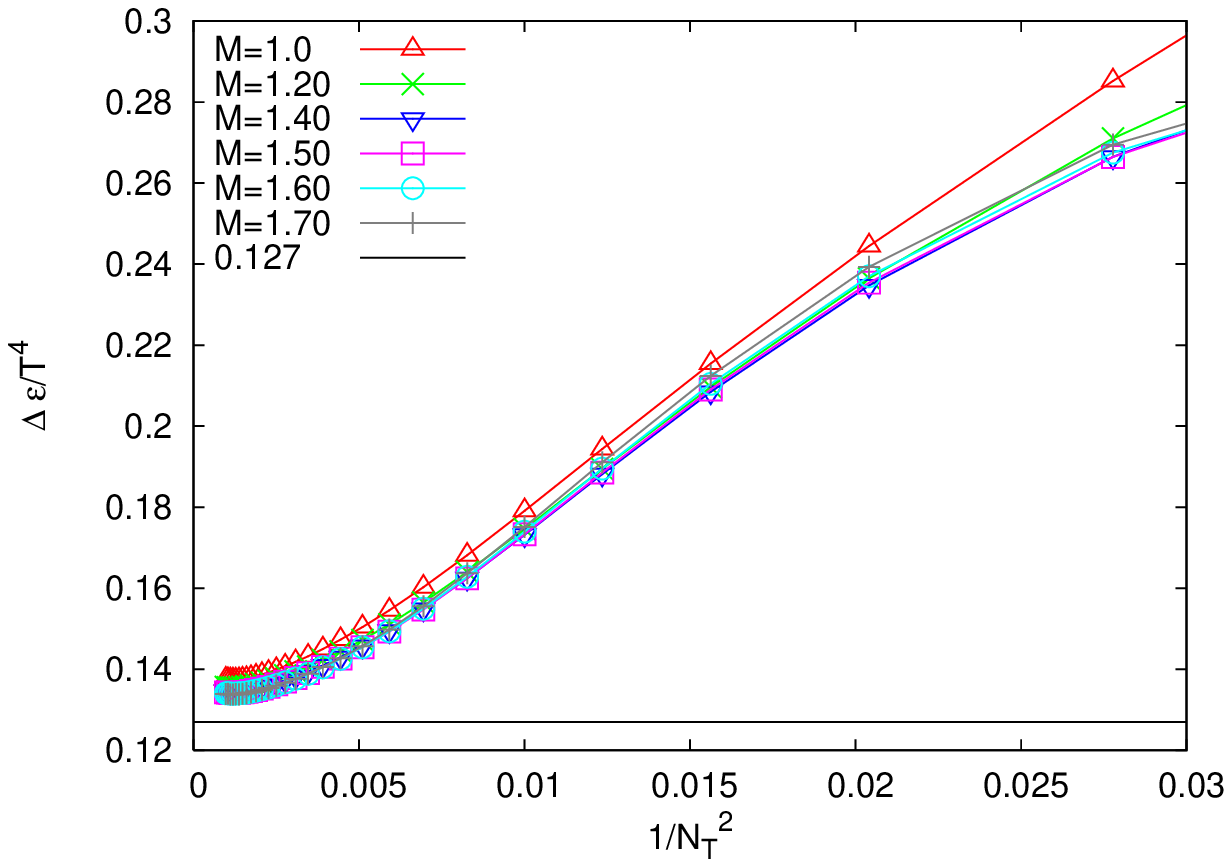}
\includegraphics[scale=0.6]{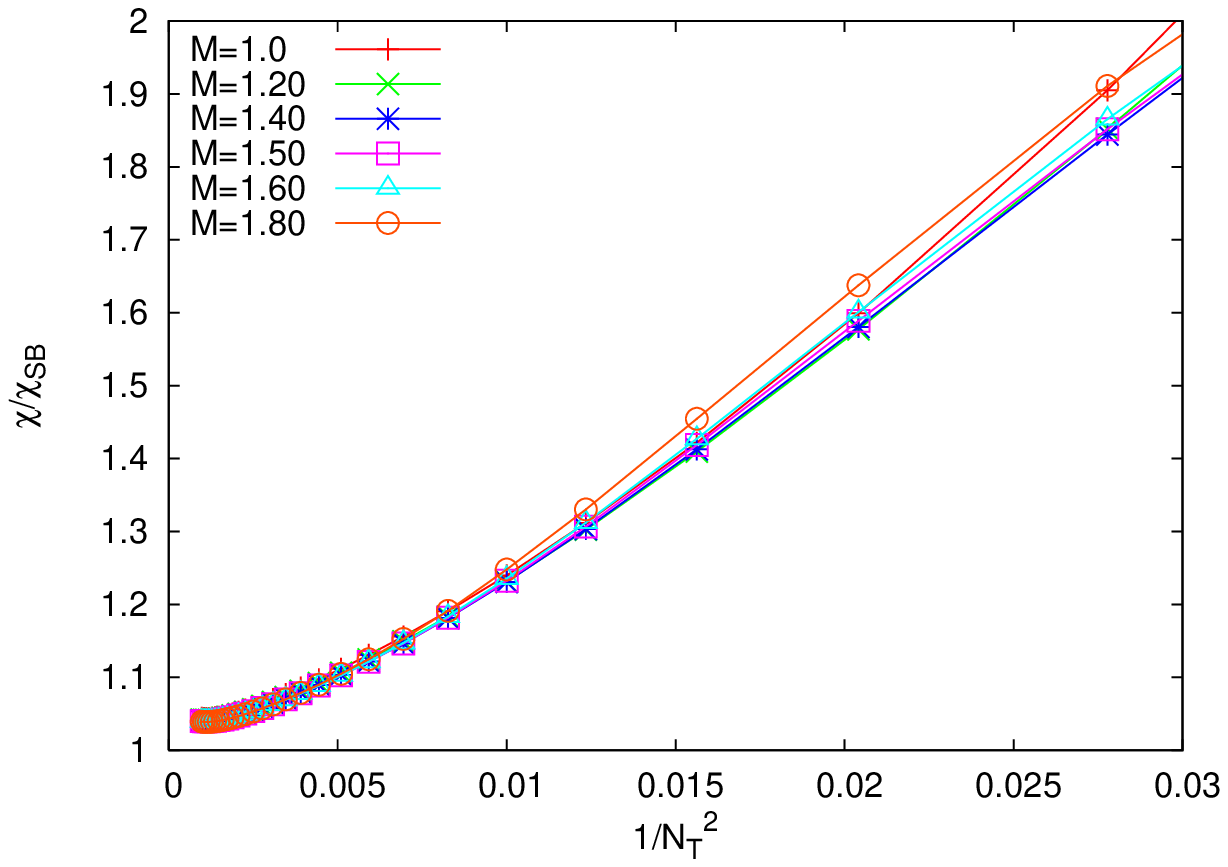}
\caption{ The energy density(left panel) in units of $T^4$ and the quark number
susceptibility (right panel), normalized to the continuum ideal gas value, as 
a function of $1/N_T^2$ for $M$ values as indicated for lattices with 
aspect ratio four. }
\label{fg:EnC}
\end{figure}

Spurred by the results of the previous sections, we propose two simple
ideas for simulations of lattice QCD at finite density.  Let us state them at
the outset: 1) If chiral invariance and the anomaly relation are to be
only restored in the continuum limit, one may introduce the chemical
potential on the lattice in a simpler way and 2) known techniques in
the literature to eliminate the free theory related divergences may
simplify the computations.

From the well-known relation between the domain wall fermions and the
overlap fermions \cite{Neu2,eh}, we know that only fermions confined
 to the domain walls are physical.  Introducing a chemical potential
 only to count them, one has
\begin{equation}
 D_{ov}(\hat\mu)_{xy}=(D_{ov})_{xy} 
-\frac{a \hat\mu}{2a_4 ~M}\left[(1-\gamma_4)U_4(y)\delta_{x,y-\hat{4}}+
(\gamma_4+1)U^{\dagger}_4(x)\delta_{x,y+\hat{4}}\right]~.
\label{eq:muN}
\end{equation}
The chief advantage of eq.(\ref{eq:muN}) is that $D_{ov}$ is defined by the
usual sign-function for a Hermitian matrix, making the computations
simpler.  Of course, it too breaks chiral invariance at the same order as
the Bloch-Wettig proposal.   One has, however, to expect
$a^{-2}$-divergences as $a \to 0$.  Following the same prescription which
is used for the pressure computation (which diverges at zero temperature as
$\Lambda^4$ ), we use physical quantities computed on large $N_T$ and
the same lattice spacing $a$ for subtraction of these divergences.  We tested
this by considering two physical quantities for the ideal gas, namely, the
difference in the energy density due to nonzero $\mu$, defined by $\Delta 
\epsilon (\mu,T) = \epsilon(\mu,T) - \epsilon(0,T)$ and  the quark number 
susceptibility, $ \chi = T/V ~ \partial^2 \ln {\cal Z} / \partial \mu^2$. 

Figure \ref{fg:EnC} shows the results for these quantities, where we used 
$\mu/T = 0.5$ for the $\Delta\epsilon $ and $\mu=0$ for the $\chi$
 to compare with
our earlier \cite{bgs} results using the exponential prescription of 
Bloch-Wettig.  The zero temperature values were computed
numerically on a lattice with a very large temporal extent $N_T$ and
fixed $a_4$ such that $T=1/(N_T a_4)\rightarrow0$.  The Matsubara
frequencies then become continuous and hence could be integrated upon
numerically. The $M$-values are as indicated along the symbol used.
From a comparison of the plots with the
corresponding ones \cite{bgs} for the Bloch-Wettig case, we find that
i) the zero temperature subtraction procedure does indeed eliminate
the divergences, ii) there are no oscillations for odd-even values of $N_T$,
iii)the M-dependence is much less pronounced, and
iv) the scaling towards the continuum value is linear with the
possibility of an easier extrapolation.

\begin{figure}
\begin{center}
\includegraphics[scale=0.65]{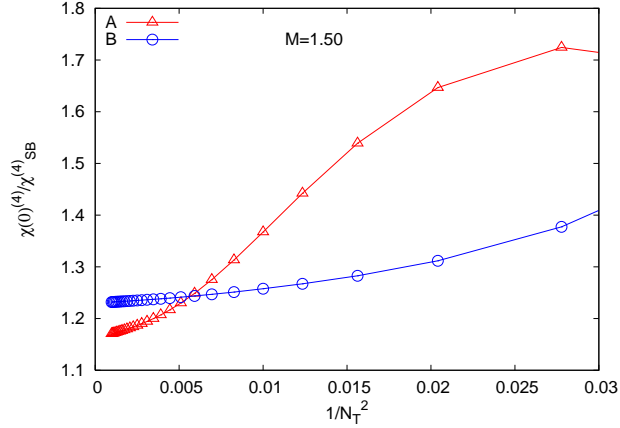}
\caption{ The variation of fourth order susceptibility, normalized by its
corresponding continuum value, as a function of $1/N_T^2$ for $\zeta
=4,~M=1.5$ for A) $\hat\mu/s$ and B)$\hat\mu/M$ way of incorporating the
chemical potential.}
\label{fg:sus4}
\end{center}
\end{figure}

A further test of the subtraction procedure is that no {\em additional}
divergences should be seen in other quantities as well, e.g., the fourth
order susceptibility. Figure \ref{fg:sus4} demonstrates this to be the
case. The two curves shown are different possible ways of normalizing
the chemical potential term, as indicated. Further improvements in the form
to achieve a faster convergence to the continuum needs to be explored.

Taking a cue from eq.(\ref{eq:muN}), one can introduce  $\mu$ in general by 
\begin{equation}
S_F = \sum_{x,y} \bar \Psi (x) M(\mu;x,y) \Psi(y) 
    = \sum_{x,y} \bar \Psi (x) D(x,y) \Psi(y) 
    + \mu a \sum_{x,y}  N(x,y)~~,
\label{eq:muNS}
\end{equation}
where $D$ can be the the staggered, the Wilson-Dirac or any other suitable
fermion operator and $N(x,y)$ is the merely the corresponding point-split
and gauge invariant number density.  Clearly, as above, one ought to be
able to get rid of the annoying divergences by subtracting the same
physical quantity computed on large $N_T$ and the same lattice spacing $a$.
Usually, one would introduce exp($\pm a \mu$) factors in $M$ and not add
the $N$ term. Denoting by superscripted primes various derivatives with
respect to $a \mu$, one sees that the usual case has $M' = M'''...=
\sum_{x,y}N(x,y)$ and $ M'' = M'''' = M''''''...\ne 0$, i.e, {\em all}
derivative terms contribute making the successive Taylor coefficients have
more and more terms. In contrast, eq.(\ref{eq:muNS}) leads to $M' =
\sum_{x,y} N(x,y)$, and $ M'' = M''' = M''''... = 0$, i.e., a lot fewer
terms in the Taylor coefficients. This could potentially help in obtaining
higher order coefficients than computed so far.  For example, the 4th (8th)
order susceptibility, has 4th (8th) derivative of $M$, which has only one
(one) term for eq. (\ref{eq:muNS}) in contrast to five (eighteen!) in the
usual case \cite{gg}.  Whether the subsequent reduction in the computations
of $M^{-1}$ actually helps in speeding these computations is currently
being investigated by us.

\section{Summary}

We showed, both perturbatively and non-perturbatively, that the
introduction of nonzero chemical potential, $\mu$,  leaves the anomaly
unaffected.  The zero modes of the Dirac operator for $\mu=0$ govern it.
Nonzero $\mu$ simply scales the eigenvectors, makes the right and left
eigenvectors distinct but together they still satisfy a completeness
relation.  We pointed out further that the reason for this can be traced to
a ``gauge-like'' symmetry in the continuum using which $\mu$-dependent
terms can be ``gauged'' away.  All the currently known prescriptions of
introducing $\mu$ on lattice can be understood in terms of a similar
symmetry on the lattice, and ought to protect the anomaly if the fermion
regularization itself has it.  While such a symmetry can be shown to exist
for local actions, it could be simply adopted for the nonlocal cases such
as the overlap fermions.  The resultant effective overlap operator for
nonzero $\mu$ also satisfies the Ginsparg-Wilson relation but unfortunately
is not invariant under the chiral transformation.

We also showed that investigating the overlap fermions at finite density
by simply adding the $\mu$-term linearly is feasible.  It too has similar
chiral symmetry breaking as the Bloch-Wettig proposal but the corresponding
inverse propagator is simpler and better-defined.  Extending the idea of
addition of $\mu$ linearly to the usual staggered fermion case was shown to
open a possible avenue for computations of higher order coefficients.

\end{document}